# IEC 61499 vs. 61131: A Comparison Based on Misperceptions

Kleanthis Thramboulidis
*Electrical and Computer Engineering*
*University of Patras, Greece*
thrambo@ece.upatras.gr

**Abstract.** IEC 61131 has been widely accepted in the industrial automation domain. However, it is claimed that the standard does not address today the new requirements of complex industrial systems, which include among others, portability, interoperability, increased reusability and distribution. To address these restrictions, IEC has initiated the task of developing IEC 61499, which is presented as a mature technology to enable intelligent automation in various domains. This standard was not accepted by industry even though it is highly promoted by the academic community. In this paper, a comparison between the two standards is presented. We argue that IEC 61499 has been promoted by academy based on unsubstantiated claims on its main features, i.e., reusability, portability, interoperability, event-driven execution. A number of misperceptions are presented and discussed. Based on this, it is claimed that IEC 61499 does not provide a solid framework for the next generation of industrial automation systems.

## 1. Introduction

Industrial automation systems have been based for many years on the IEC 61131 standard [1], which was first published in 1992. The standard, which is considered as one of the most important ones in industrial automation [2], defines a model and a set of programming languages (part 3) for the development of industrial automation software [3]. Control engineers predominantly use it to specify the software part of their systems, mainly when programmable logic controllers (PLCs) are used [4]. However, the standard has been criticized the past few years for not addressing any more the requirements of today's complex industrial automation systems and not being compliant with state of the art software engineering practices. For example, it is claimed in [5] that current software architectures of industrial process measurement and control systems (IPMCS), such as IEC 61131-3, do not conceptually support reconfiguration and distribution. In addition, as claimed in [6], the concepts of 61131 "are not the state of the art in software engineering anymore", even though not explicitly stating which concepts the standard does not support and which of the supported ones are not any more state of the art. On the other side, portability, configurability, interoperability, reconfiguration, and distribution have been identified in [5] as the high-level demands/requirements for future automation systems.

   To address restrictions as well as new challenges in the development of industrial automation systems, the Technical Committee 65 of the International Electrotechnical Commission (IEC TC65) was assigned the task of developing a new standard. IEC 61499 [7], which is the result of this work, has been promoted in the literature as the solution that would address restrictions and challenges in the industrial automation development process. For example, 61499 is presented in [5] with the following main features: "component-oriented building blocks called FBs, graphical intuitive way of modeling control algorithms through the connection of FBs, direct support for distribution, interaction between devices of different vendors, and basic support for reconfiguration, and it is based on existing domain standards." Moreover, the IEC 61499 is considered to have emerged in response "to the technological limitations encountered in the currently dominating standard IEC 61131" [8]. It appears as a new programming paradigm that is going to replace the IEC 61131, which is already established and widely used in industry. A significant number of publications in prestigious journals promote the use of 61499 as an enabler of distributed and intelligent automation, e.g., [9], or multi-agent control software architectures, e.g., [10]. Modern software engineering methodologies have been adapted, according to [6], with IEC 61499, to the domain of industrial automation. More specifically it is claimed that requirements such as flexibility, adaptability, or robustness, which are not successfully addressed by existing design methods, are successfully fulfilled by the IEC 61499.

However, since "industrial deployments are still in the early phases of experiments and feasibility studies" [4], IEC 61499 was not adopted by industry. This is admitted even by its supporters. For example in [6], it is admitted that "the majority of control device vendors and users (e.g., control engineers) still cannot appreciate the advantages of using IEC 61499", but this is ascribed to the "many misconceptions of the standard's ideas and contradicting conclusions of the researchers on the role and place of the standard in future industrial automation." Misconceptions are based, according to [6] on one of the first books on the IEC 61499, i.e., [11], which is based on the first draft version of the standard. In the same paper, even though it is recognized that there is a lack of comprehensive case studies comparing the overall benefits and drawbacks of both technologies, several drawbacks of the IEC61131 are presented based on bird's eye view and IEC61499 is promoted as a new paradigm that addresses these drawbacks. Moreover, the claim that IEC 61499 "combines several technologies, targeting portability, configurability and interoperability of automation systems" [9], is not substantiated.

On the other side, IEC 61499 is highly criticized, e.g., [12][13][14], for its inability to address its objectives, i.e., the challenges in industrial automation systems development. Authors in [15] argue that "the standard is ambiguous and different implementations of the standard have made different assumptions to cope with the ambiguities." This has as result, according to [15], the same application to behave differently when it is executed on different implementations, "thus making the applications unportable." Moreover, even its strong supporters, including the task force leader of the corresponding working group of IEC 61499, have admitted that "even stricter and more precise provisions are required in order to achieve the main goals of the IEC 61499 standard that are portability, (re)configurability, interoperability, and distribution" [5].

In this paper, we focus on the arguments used in the IEC 61499 literature and we present a number of misperceptions that are used in its comparisons with IEC 61131. We claim that the positives of IEC 61499 provide minor contributions compared to the negatives introduced by the new standard. Based on this study, we claim that IEC 61499 does not provide a solid background to switch from the widely accepted IEC 61131, since 61499 introduces more problems in the development process of industrial automation systems than the ones it promises to address. In this study, the requirements context for future automation system defined in [5] and [6] is adopted. This context focuses on portability, configurability, interoperability, reconfiguration, and distribution. Our presentation is based on misperceptions that are used in past comparisons to promote the use of IEC 61499 based on restrictions of the IEC 61131. More specifically, the following misperceptions are presented and discussed in this paper.
1) A misperception related with reuse.
2) A misperception related with portability,
3) A misperception related with interoperability.
4) Three misperceptions related with the event driven nature of IEC 61499.

Other misperceptions, not addressed in this work, include among others: The object-oriented misperception, the architecture-centric development misperception, the system-level language misperception and the intelligent automation misperception.

The remainder of this paper is organized as follows: In Section 2, a brief introduction of the basic concepts of the two standards is given. In Section 3, the reuse misperception is presented and discussed. Section 4, presents and discuss the portability misconception. In Section 5, the interoperability misperception is presented. Three misperceptions related with the event driven model of IEC 61499 are discussed in Section 6 and the paper is concluded in the last Section.

**2. Basics of IEC 61131 and IEC 61499**

Fig. 1 presents the Up-Counter 61131 Function Block (FB) as defined in IEC 61131 [1]. In part a, the interface of the FB in graphical notation is shown, while in part b the body in textual format is presented. The body of FBs or programs can also be described in graphical notation using Function Block Diagram (FBD). Instances of FB types and Functions are interconnected to form FBDs to graphically specify the behavior of programs and FBs. Nodes of the FBD graph are interconnected using one type of connection which does not specify the type of information that flows from one end to the other. The direction of the flow is defined from the convention that outputs are shown on the right side of the FB while inputs on the left side. IEC 61131 supports also encapsulation of data and event driven execution of FBs and programs.

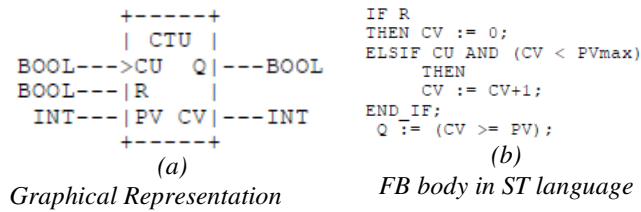

*(a)*
*Graphical Representation*

*(b)*
*FB body in ST language*

*Fig. 1.* Up-Counter 61131 Function Block [1].

IEC 61499 discriminates between event and data connections. Fig. 2 presents the same behavior with the one captured in Fig. 1, expressed as 61499 FB as defined in IEC 61499 [7]. We note that the two inputs of type BOOL, i.e., CU and R, are handled as event inputs for the 61499 FB, and two new output events are added, i.e., CUO and RO. Output events notify the environment of the FB that the corresponding behavior has been executed and the output values are available. It should be noted that the caller FB is not able to utilize the result output values of the called behavior during the current execution of the FBN. This is valid even in the case that the event connection would be characterized as synchronous and implemented using method call, as is the case of FBDK [28].

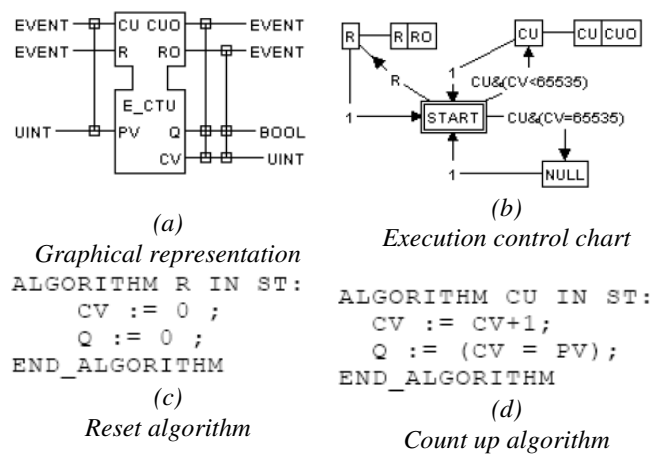

*(a)*
*Graphical representation*

*(b)*
*Execution control chart*

*(c)*
*Reset algorithm*

*(d)*
*Count up algorithm*

*Fig. 2.* Up-Counter 614199 Function Block [7].

Input events are used for the definition of Boolean expressions that trigger the transitions of the Execution control chart (ECC) of the FB, as shown in part b of Fig. 2, and "may affect the execution of one or more algorithms;" [7] . More specifically an EC transition shall have "an associated Boolean condition, equivalent to a Boolean expression utilizing one or more event input variables," Output events are issued on completion of algorithms execution.

The transformation of the behaviors captured as FBs in 61131 has not been re-organized following the OO paradigm but it has mainly done based on the separation of event and data interface provided by 61499. This is evident if we look at the string manipulation FBs of 61131 that have been converted to corresponding 61499 FBs (see FBDK) without following the OO paradigm as is the case of string management functions of C and the corresponding String class of Java.

The question today, seven years after the publication of the first version of the IEC 61499, is, if this standard has successfully researched its objectives. In this paper a comparison is presented that shows that it has failed, even though the view provided by the academic community is different.

## 3. Reuse

*Misperception 1. The reuse Misperception*

The design process in most engineering disciplines is based on component reuse, since the development of new products is based, for several reasons, on components that have been tried and tested on other systems. Reusability concerns not only small components but also major sub-systems [16]. This is also the case for software development where reuse is always a major concern since it does not only reduce the development cost but it may increase system reliability and reduce the development process risk. The benefits from the reuse of abstract products of the development process may be greater than those from the reuse of code components. Moreover, industrial automation code contains low-level detail which usually specializes it to such an extent that it cannot be reused. Design artifacts are more abstract and hence more widely applicable. Four levels of reuse are considered in [16]: application system reuse, sub-system reuse, module or object reuse and function reuse. Reuse at application level is highly related to portability, i.e., the ability of the application to be executed on several different platforms.

All the above are the reasons that reuse is considered in almost any comparison of IEC 61131 with IEC 61499. However, there are several misconceptions regarding reuse and this section attempts to bring some light into this matter. Arguments used as drawbacks for reuse on IEC 61131 include among others.
1) The support of global data that is provided by IEC 61131.
2) The absence of explicit control on the execution order of IEC 61131 FBs in an application.
3) The vendor-specific extensions.
4) The vendors' partial support of the standard.

More specifically, the first two of them. i.e., a and b, are considered in [6] as the two main drawbacks that hinder reuse in IEC 61131, even though it is recognized that the encapsulation of application parts offered by the standard helps to modularize control applications and foster the reuse of application parts. Argument b is also expressed in [6] by claiming that "the same application may work differently on different control devices". Arguments c and d are used in [6] in the authors claim that the direct reuse of developed control software elements is not really possible in IEC 61131-3 due to the vendor-specific extensions or only partial support of the standard.

Regarding the first argument, i.e., a, it is claimed that "there is no global data in IEC 61499" and that "there are no means to access or change an FB's internal variable as it was possible in IEC 61131-3 with the access path mechanism." However, it is the responsibility of the developer to effectively utilize the language constructs to obtain reusability at module, object or function level. It is widely known that the use of global data restricts reusability at these levels but it is also known that reusable objects can be developed in C++ and other languages which provide support for global data. Moreover, the public keyword is extensively used in OO languages to allow direct access to the internal variables of objects; access is also possible through accessors and mutators. These features are not an indication that these languages hinder reuse.

Regarding the second argument, it is claimed in [6] that "In an IEC 61331-3 control system, the execution order is derived from the connections between FBs, according to the rules defined in IEC 61131-3 [11, pp. 249ff.]." and also that "These rules leave some room for interpretation; therefore, the same application may work differently on different control devices."
However, it should be noted that the 61131 languages allow for the explicit definition of the execution order of FB instances, with the only exception being the graphical representation of the FBD. But even in this case, all commercially available tools address this problem by adopting a notation that allows the developer to explicitly specify the execution order. The problem can be addressed in the graphical representation in a similar way with the one adopted in the graphical representation of C language expressions to explicitly define the evaluation order (see abstract syntax tree). Moreover, it is not mentioned that the IEC 61499 standard completely fails in defining execution semantics and that the existing IEC 61499 execution environments, commercial and academic, have completely different execution semantics, making the reuse of any 61499 artifact impossible on other execution environment. This is reported in [15] where it is claimed that the same application behaves differently when it is executed on different implementations, "thus making the applications unportable." This is also claimed in [8], according to which the IEC 61499 standard "does not provide formal semantics for the execution of function blocks" but instead it contains "a verbose description for function block execution, which has resulted in multiple interpretations." In the same paper, it is also recognized that the same design may behave "differently when executed on the various existing runtime environments, such as FBRT [4], RTSJ-AXE [5], FORTE [6], Fuber [7], and ISaGRAF [8]."

Finally, regarding the last two arguments, i.e., c and d, it is clear that no one may prevent the IEC 61499 vendors to implement vendor specific extensions or partially support the standard, as is the case today with all the execution environments. So, these two arguments cannot be seriously considered in a comparison.

From all the above it is evident that there are no serious arguments on the advantage of 61499 compared to 61131 regarding reusability. Moreover, it should be noted that the upcoming version of 61131 with its object oriented extension provides better mechanisms for reusability, through the constructs of class and interface and the keywords extends and implements [17].

**4. Portability**

*Misperception 2. The portability Misperception*

Portability is defined as the ability of a system to run under different computing environments, hardware and/or software. In other words it is the ease with which behavior can be moved from one system to another. This means that portability is a prerequisite for reuse across different computing environments. We discriminate between portability at source code level (edit-time) and portability of executable code level (run-time). For example, bytecodes in Java provide support for executable level portability.

Several journal publications present the IEC 61499 standard as enabling portability in the industrial automation domain. For example, portability is considered in [18] as the first benefit that comes from the use of IEC 61499 to the deployment to distributed embedded targets. And FBs are considered to be "portable and have platform-independent execution semantics." It is also claimed that the standard "provides a portable high-level executable specification framework for distributed automation." The same is claimed in [19] where IEC 61499 appears to offer a number of "flexibility provisions for distributed control system which are not available in IEC 61131-3 PLCs." Among them, portability of IEC 61499 function block-based design is mentioned as a convenient way to reuse functionalities of nodes. In [20] it is claimed that IEC61499 "enables encapsulation, portability, interoperability, and configurability." In [21], it is claimed that it is not easy to have an IEC 61131 program to "run on a PLC of some other vendor", while for IEC 61499 it is claimed that it offers "several semantic and syntactic provisions for portability." The use of XML to specify the design artifacts is presented, in the same paper, as the mean on the syntactic level "which enables easy exchange of source files between tools of different vendors, as well as interaction of tools with devices of different vendors."

We claim that the feature of portability of IEC 61499 is unsustainable. Regarding the arguments presented in [21], it is not explained how the features of the standard, which are mentioned as semantic provision of IEC 61499 towards portability i.e., encapsulation of data, the event driven nature of FBs and their composability, influence portability. Moreover, regarding the mentioned semantic provisions, it should be noted that IEC 61131 supports encapsulation of data, event driven execution of FBs and programs. It also supports composability in a similar way with IEC 61499 since a function block network can be used to specify the body of a Program or FB [17]. Furthermore, PLCopen XML allows the generation of portable designs based on 61131 and this is already supported by commercial tools. On the other side, there are no two IEC 61499 tools available in the market that are able to exchange 61499 designs, preserving the same semantics. This is also admitted in [6] which accepts that the weak semantic-related descriptions in IEC 61499-1, have been interpreted differently by different execution environments. This has resulted, according to the authors, in having the same application to behave differently on different execution environments. And this was the main reason that O3neida is currently developing a compliance profile defining the execution behavior of 61499 run-time environments. However, this cannot be used as an argument to claim that 61499 supports portability.

The absence of formal semantics for the execution of function blocks is also reported in [8], where the authors claim that the standard contains a verbose description for function block execution, which has resulted in multiple interpretations. Moreover, author admit that this "hampers portability (defeating the purpose of a standard)." As authors claim in [5] "stricter and more precise provisions are required in order to achieve the main goals of the IEC 61499 standard" among which portability. Moreover, portability is one of the benefits of IEC 61499, that has not been the major concern of the vendors yet, even though this feature is ""embedded" and hopefully will be recognized in a short time span", as claimed in [9]. In the same paper, it is also claimed that the IEC 61499 combines several technologies targeting portability, without mentioning these technologies. Portability is also considered as the main motivation for the event-driven execution, that is characterized as "the desire to make the code independent of the sequence of FB invocation in the PLC scan loop." Strong data encapsulation into components is considered as another provision for portability.

From the above, it is evident that portability cannot be considered as advantage in the comparison of IEC 61499 with IEC 61131.

## 5. Interoperability

*Misperception 3. The interoperability Misperception*

Interoperability in its simple definition is the ability of two or more systems to cooperate at runtime. Based on [22] the following broader definition can be given for industrial automation systems. Interoperability is defined as the ability of heterogeneous devices (PLCs) to interact towards mutually beneficial and agreed common goals, involving the sharing of information and knowledge between them, through the behavior they support, by means of the exchange of data.

As claimed in [19], IEC 61499 "aims at the flexibility, interoperability, portability, and reconfigurability which are not sufficiently supported in the IEC 61131-3 standard" and in this way it "provides the next generation systems' architecture enabling distributed control in automation industry." The only argument used for interoperability is that "IEC 61499 Function Blocks are stored in an open XML-based format" which "allows for better interoperability between platforms as well as for greater reusability." And the conclusion regarding interoperability is that "Interoperability of IEC 61499 devices facilitate the autonomous reasoning, distributed decision making and negotiation between nodes."

The use of XML is also mentioned as a mean for achieving interoperability in [9] but this solution is considered as "not appropriate for many embedded platforms" since "XML tools are quite performance hungry." It is also mentioned that a solution to this problem has already proposed based on binary XML but this solution "impacts on interoperability, as there are several different versions of binary XML, supported by different user groups" [9].

In [20] it is claimed that "IEC 61499-compliant devices can easily interface with each other, thus providing for seamless distribution of different tasks across different devices" and since the user "may create his/her own program using standard *FB* types", the IEC61499 architecture "enables encapsulation, portability, interoperability, and configurability." The definition of interoperability is also given as the ability of hardware devices to "operate together to perform the cooperative functions specified by one or more distributed applications." However, it should be noted that it is admitted, at the end, that "the industry standards necessary for universal interoperability and cost reduction are very early works, which are in progress."

In [2] a detailed discussion on interoperability issues regarding 61131 is given. Regarding IEC 61499, it is claimed that it "offers the possibility to provide highly integrated solutions with devices of multiple vendors" though the mechanism of compliance profiles." It is claimed that these profiles, once defined, will allow any differentiation, but these differentiations will be documented and consistently defined in the corresponding compliance profiles. In a similar way, it is admitted in [6] that the standard does not address the interoperability requirement since it is quite natural that "the standard cannot foresee upfront all the features of devices' programming, configuration, or communication that need to be standardized. Instead, it defines a flexible and extensible mechanism of compliance profiles" and it proposes the definition of profiles to address interoperability. It should be noted that, as admitted in [2], the practice up to now "showed that the so advantageous sounding properties portability, configurability, and interoperability are also not given a priori with IEC 61499." Moreover, and since "efforts toward portability, configurability, and interoperability need the common work and specification of a broad positioned user organization", O3neida has "dedicated itself as discussion platform" to achieve the goals of the standard.

In [9] it is claimed that "The IEC 61499 standard combines several technologies, targeting portability, configurability and interoperability of automation systems", even though it is admitted that "portability and interoperability have not been the major concern of the vendors yet" but as claimed "these features are "embedded" and hopefully will be recognized in a short time span."

In [5] it is claimed that even though 61131-3 provides a small basis for common modeling of control programs, platforms and tools are not able to interoperate. Regarding 61499, it is claimed that the investigation on interoperability showed open points in the definitions of the standard for application modeling when executed on different devices and that "even stricter and more precise provisions are required" in order to achieve the goal of interoperability.

Currently there are no two IEC61499 execution environments that support interoperability in the context of a distributed application. From all the above, it is evident that IEC 61499 does not provide support for interoperability and there are no significant arguments to prove it. A proposal to obtain interoperability in IEC 61499 based systems is described in [23] using CORBA.

## 6. The Event Driven nature of IEC 61499

Two different approaches to the design of the control mechanisms of real-time computer applications can be identified, depending on the type of triggering mechanism used [24]:
   a) time triggered control, and
   b) event-triggered control.

In time-triggered control, all activities are initiated by the progression of real-time. In event-triggered control, all communication and processing activities, are initiated whenever a significant event other than the regular event of a clock tick occurs. Both approaches can be used to address a control problem assuming their correct realization. As claimed in [13], "the same behavior should also be ensured for the case of an event-based implementation or a cycle-based one", i.e., the design specification of FB-based applications should be independent of the implementation method used to interface with the controlled process. The suitability of the IEC 61499 model for event-driven control at the physical device level of the shop floor is considered in [25] as the primary motivation for its use. However, there are several misperceptions in the IEC 61499 literature regarding the meaning of event driven execution.

*Misperception 4: Event-driven vs. scan based*

There is a misperception regarding the relationship of the terms event-driven and scan based execution. Scan based execution is a term used to refer to time-triggered control that is widely applied in 61131 based systems. For the IEC 61499 it is not clear if the event driven nature refers to the handling of external events or to the handling of the internal ones. Most of the IEC 61499 journal papers consider it as referring to the handling of the internal events. Others relate it to the handling of externals. Both categories examine the relation or mapping of the event driven to the scan based.

Authors in [2] claim that "Choosing the event-driven execution model of IEC 61499 or the cyclic execution model of IEC 61131 is application-dependent." So, the event driven nature of IEC 61499 appears to offer a solution, or to be an alternative to the cyclic scan model widely used in 61131.

Authors in [5] refer to a mapping from the event-based IEC 61499 execution to a "cyclic, scan-based" system as it is used in traditional programmable logic controller (PLC) systems. The authors claim that: a) this mapping allows a deterministic preverifiable execution behavior and response times, and b) that the large overhead introduced by the scan cycle is the main limiting point in scan based systems and it results in larger overall response time for the application. The authors also cite a paper that compares an event-based implementation of IEC 61499 with the scan-based execution model, which has been implemented within the commercial tool ISaGRAF, and shows that "events maybe lost in the event-based execution model."

In practice, scan-based is nothing more than an implementation of polling. But this is also the case of the 61499 FBN shown in Fig. 3, which is referred in the 61499 community as event based. So, it is clear that regarding this subject the only difference is that the behavior is executed, in the 61131 case, by the thread of the task, while in the case of 61499 the behavior is executed by the thread of TankFB. But this later, raises the question of where to assign the properties of this thread. The solution to define them in the TankFB reduces its reusability since the decision to implement it using periodic or triggered execution is captured in the FB. The 61131 approach sounds more modular and preserves the application design model from implementation related issues.

*Misperception 5: IEC 61131 does not support event triggered control*

There is a misperception created by many conference and journal publications that the IEC 61131 does not support event-triggered control. As claimed in [8] "the IEC 61499 standard has emerged in response to the technological limitations encountered in the currently dominating standard IEC 61131", with one of these being the cyclic scan model of computation, which is considered as "severely inadequate to meet the current industry demands for distributed, flexible automation systems." It is also claimed that in order to meet challenges, among which the above, the IEC 61499 "has defined a new event-driven model for function blocks intended for distributed execution." In [5], it is claimed that the IEC 61499 "extends the FB model of its predecessor IEC 61131-3 with additional event handling mechanisms and concepts for distribution." In a similar way, it is claimed in [26], that the IEC 61499 was initiated as an extension of IEC 61131-3 Function Blocks by adding event driven execution, to address limitations of the legacy PLC programming languages.

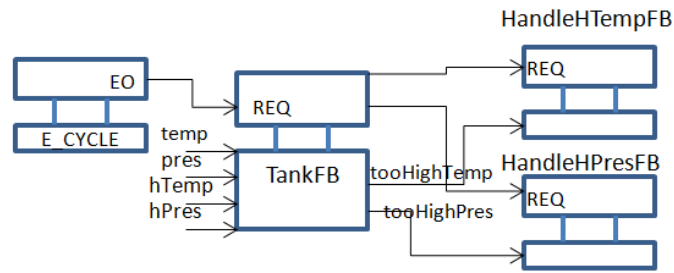

*Fig. 3*. IEC 61499 Tank FB application with E_CYCLE.

However, the IEC 61131 standard supports both periodic and triggered execution of tasks and so the periodic and triggered execution of the group of associated with the task programs or FBs. Scan-based is a term used to refer to the use of periodic tasks for the execution of the POUs that capture the behavior of the application to an event or a set of events. In particular, Part 1, that is entitled 'General Information', refers to two execution control models, i.e., periodic and event-driven execution of "a series of instructions stored in application programme storage." Part 3, that is entitled 'Programming Languages', defines the task as an execution control element which is capable of calling, either on a periodic basis or upon the occurrence of the rising edge of a specified Boolean variable, a behavior expressed in the form of a program or a function block.

*Misperception 6:* *The event driven execution Misperception*
It is consider that the objective of the event interface is the explicit definition of the execution sequence of FB instances in a Function Block Network (FBN). For example in [27], the change from the "data-driven approach" of IEC 61131 to the "event-driven approach" of 61499 is considered as the second major change introduced by IEC 61499. It is claimed that this offers the advantage to the control engineer to explicitly specify the order of execution in IEC 61499 applications though the additional event connections. In a similar way, it is claimed in [6] that the event interface "allows for explicit specification of the FBs' execution sequence" resulting to "a new level of flexibility" for the developer, that is not offered in IEC 61131-3. As a result of this, 61131 FBs performing string manipulations have been converted to 61499 FBs having REQ and CNF events, as is the case with the string concatenation FB shown in Fig. 4 [28].

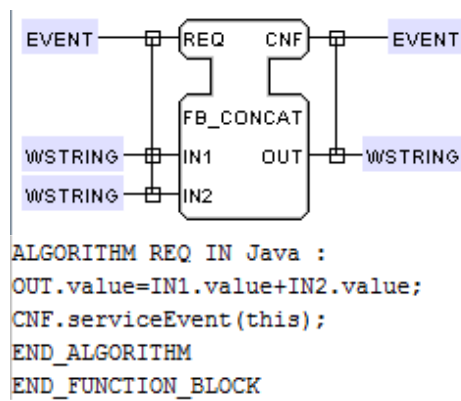

*Fig. 4*. FB for concatenation of WSTRING inputs [28].

In [9], it is claimed that the implementation of an event-driven activation of function blocks "implies the possible need of storing events in queues of a variable length and loss of events in case the queue capacity is exceeded." This may lead, according to the author, to non deterministic behavior that has been addressed by the IEC 61499 community by several proposals such as the synchronous model of execution, the cyclic model of execution, and the ISaGRAF model, which is considered as close to the cyclic. An "application-level practical solution" is proposed towards this direction, to achieve the determinism of event-driven applications. According to this, external inputs are sampled periodically, using the E_CYCLE FB, and delivered to the rest of the application if any change is detected. This solution is presented as "preserving the event-driven nature of the application." The author claims that this solution is different from the PLC scan, since "the controller application executes only when a change of any input is detected." Moreover, another benefit, according to [9], is that "the event-driven execution also can help in activating only those blocks which are directly dependent on that event, unlike the PLC case, where all program modules are executed, which may result in "less time for execution even in the worst case." In the same paper, FBRT is referred as a "pure" event-driven implementation platform, even though it is well known that the FBRT implements the event connections between FB instances using method call. As claimed in [5] "the event-notification mechanism is implemented as a simple function call."

Moreover, IEC 61499 does not adopt an event triggered approach to handle the external events. As claimed in [5] "Up to now, there exist no mechanisms to introduce any events into applications, which is a critical issue. The responder SIFBs are so-called resource initiated and they are special SIFBs that are triggered by an external event source (ES). Therefore, such specific SIFBs generate events for applications. This means that an underlying service (e.g., timing interrupt or network invocation) activates the execution of FBs." It also states that external event sources "(e.g., timers) are mapped to own Java threads."

To analyze the misperception on the event driven execution, we use as an example the FBN presented in [21] for the control of a pneumatic cylinder, that is shown in Fig. 5. The same example is used in [29]. BUTTONS and LIGHT are implementing the, what authors call, "resource-initiated service model", which means that "upon any change of the source signal, e.g., light curtain status, the corresponding FB is activated without any input event (…)". POSITION FB, on the other side, is activated periodically by the pulse generator E_CYCLE FB. However, it is not clear what is the nature of BUTTONS and LIGHT FBs. It is mentioned that they are activated by the resource but it is not known if the resource acts in a time-triggered or event-triggered mode.

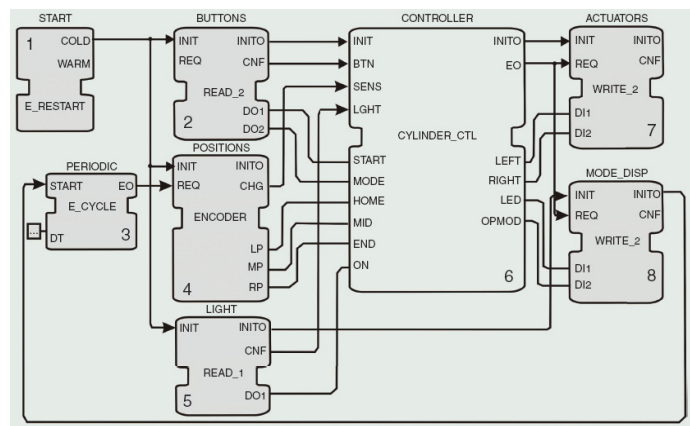

*Fig. 5.* FBN with time-triggered and event-triggered control [21].

It should be noted that with a notation like this, FBs have hidden interfaces and dependencies, which are not shown in the design specification, i.e., external input and output events are not shown on the application's design specification. Furthermore, it is questionable if the behavior to these dependencies is captured in their corresponding ECCs.

It is also interesting to discuss the ECC of the CYLINDER_CTL FB, which is shown in Fig. 6, as given in [29]. Even though the semantics of this state machine are not known, so it is unreadable, one may comment on the difference on handling the INIT event from the other events. There is a specific state, i.e., the WAIT one, to model the time the state machine is waiting for an event input excluding INIT. The wait behavior is not modeled as state in other cases, e.g., in states START, STOP or ZONE0 and ZONE1. Moreover there are several states, e.g., SELECT, that represent nothing. But the most important is that this state machine handles responses to the SENS periodically activated event and also to events resulted from interrupts, as claimed by authors, i.e., LIGHT and BTN. And of course it is not obvious what is the behavior in case an interrupt occurs during the processing of the SENS event or what is the behavior in case the LIGHT interrupt occurs when the state machine is in state ZONE0.

Moreover, it should be noted that even though the authors use the E_CYCLE FB to implement in practice the cyclic scan model of 61131, they consider and examine on the same FBN the application of the cyclic scan model without excluding the E_CYCLE FB.

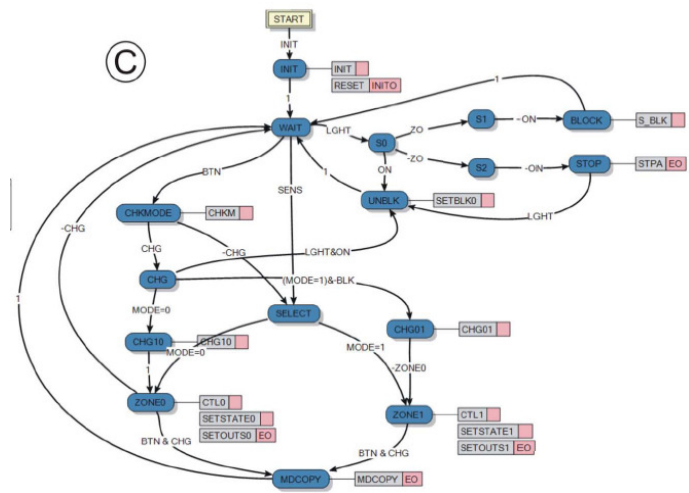

*Fig. 6.* Execution Control Chart (ECC) for an event and time triggered FB [21].

We assume that there is a need for the body of the cylinder FB to perform a comparison of two reals, i.e., there is a need to use the functionality offered by the FB_LT_REAL FB, shown in Fig. 7, as defined in the FB library of the FBDK [28]. This FB encapsulates just a function; the REQ input event is used to trigger the FB, which, once triggered, executes the comparison and fires the CNF event output, while the result of the comparison is stored in the OUT data output. Amazingly there is no way to show in the FBN such a call of behavior. This seems to work, if FBDK is used, due to the implementation of the event connections as method calls and assuming that the CNF will not be fired. However, even though the behavior captured by the algorithm of the body will be executed and the result of the comparison will be stored to the data output OUT and the control will be transferred to the cylinder's state machine, since there is no event fired, the result would not be used, even if the corresponding output is connected to an input of the cylinder FB. This is due to the semantics of the ECC, according to which data inputs are read once when the ECC is activated. If an event based implementation of event connection is assumed, it is not possible to implement this very simple behavior call.

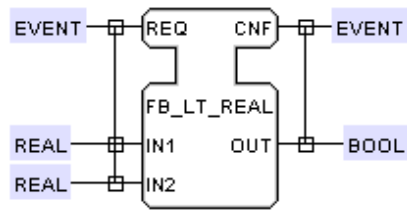

*Fig. 7.* An IEC 61499 FB indicating the misperception on the event-driven execution.

The misperception about the event driven model is evident if we consider the IEC 61499 Compliance Profile for Execution models, that defines according to [9], an execution model. The objective of this profile, according to [30], which briefly describes it, is "to establish a draft compliance profile [15] for varying IEC61499 architectures and establish a means of ensuring compatibility in execution for future implementations of the standard". Section 5.1 of this profile, identifies the need to "have a predefined order before execution, such that during the course of a scan, all FB's within an FBN will always be invoked in the same order"[30]. This is obtained by having the designer to assign a unique priority to each FB of the FBN. This priority is actually defining the execution order of the FB instances during every scan cycle, i.e., "the tool invokes all the FB's within a resource's contained FBN in a fixed order (schedule)"[30].

ISaGRAF combines, according to [30], "the scan-based logic execution of the IEC 61131 PLC standard, with the EDI concept of 61499, known as the "Cyclic Execution Semantics", where EDI stands for event-driven invocation. In addition, as claimed in [6], the "event driven IEC 61499 FBs are executed on top of a cyclic scanned and time-triggered IEC 61131-3 run-time system."

From the above it is clear that the execution model proposed by the Compliance Profile is based on the key concepts of the execution model of the IEC 61131 FBD. This is in antithesis with the main objective of IEC 61499 and this is captured very well in [8], where it is claimed that the IEC 61499 standard has emerged in response to the technological limitations encountered in the currently dominating standard IEC 61131, with one of these being the cyclic scan model. In the same paper the 61131 model of computation is criticized as "severely inadequate to meet the current industry demands for distributed, flexible automation systems."

## 7. Summary

IEC 61499 has been proposed by IEC to address the restrictions imposed by the IEC 61131 standard in the development of today's demanding industrial applications. The standard was promoted by the academic community as addressing requirements among which reusability, portability, interoperability and distribution, which are imposed by today's complex industrial systems. Several journal publications exploit these features of the IEC 61499 to propose it as new more robust and effective platform for industrial applications compared with IEC 61131. However, as claimed in this paper, the advantages related with these features are unsustainable on the comparisons with IEC 61131. A number of misconceptions on these features are presented and discussed to show that IEC 614499 does not define a platform that provides considerable support for these features, i.e. reusability, portability, interoperability and distribution. Arguments used to show that IEC 61499 provides better support for reuse, portability and interoperability are not existing or very weak and the results of the comparisons unsustainable. The event driven model of IEC 61499, which is considered as enabling technology towards these features is discussed in detail to identify and present three misperceptions related with it. It has been shown that 61499 cannot be considered as the effective successor of 61131 not even provide, at least with the current version, a reliable alternative for the development of industrial automation systems.